# Theory of multiexciton generation in semiconductor nanocrystals

Eran Rabani[(a)] and Roi Baer[(b)]

*(a) School of Chemistry, The Raymond and Beverly Sackler Faculty of Exact Sciences, Tel Aviv University, Tel Aviv 69978 Israel; (b) Institute of Chemistry and the Fritz Haber Center for Molecular Dynamics, The Hebrew University of Jerusalem, Jerusalem 91904 Israel.*

Draft: Wednesday, July 07, 2010

We develop a generalized framework based on a Green's function formalism to calculate the efficiency of multiexciton generation in nanocrystal quantum dots. The direct/indirect absorption and coherent/incoherent impact ionization mechanisms, often used to describe multiexciton generation in nanocrystals, are reviewed and rederived from the unified theory as certain approximations. In addition, two new limits are described systematically - the weak Coulomb coupling limit and the semi-wide band limit. We show that the description of multiexciton generation in nanocrystals can be described as incoherent process and we discuss the scaling of multiexciton generation with respect to the photon energy and nanocrystal size. Illustrations are given for three prototype systems: CdSe, InAs and silicon quantum dots.

## I.   INTRODUCTION

The development of efficient and cheap devices that utilize solar energy is one of the grand challenges in modern science (1). In recent years, much attention has been given to the development of light-harvesting devices based on nanostructured thin-film materials(2-5). These materials offer the promise of low cost, small dimensions, light weight, and efficiencies up to the Shockley-Queisser (SQ) limit of 31% for single junction devices (6).

While reaching the SQ limit still remains a challenge for thin film nanostructured technology, there exist several concepts that hold the potential to move efficiencies beyond the SQ limit, to as much as 66% (7,8). One approach, which will be covered in the present work, is based on the generation of multiple pairs of charge carriers from a single absorption event. This process has been referred to as "Multiexciton Generation" (MEG) which can lead to "Carriers Multiplication" (CM) (7).

The key idea behind the generation of multiexciton upon the absorption of one photon is sketched in Figure 1: for the special case of generating a biexciton. The absorbed photon creates an exciton composed of two charge carriers: A negative electron and a positive hole, each having an effective mass depending on the band structure of the nanomaterial. The exciton can either decay, typically by phonon emission, to the band edge with a timescale of $\hbar\gamma^{-1}$ ($\gamma$ is also the imaginary part of the phonon self energy, see below) (9). The competing process, which is the one of interest in the present work, is the transformation of the excitonic state into a resonant biexcitonic state with a timescale $\hbar\Gamma_S^{-1}$ ($\Gamma_S$ is also the imaginary part of the biexciton self-energy, see below). This biexcitonic state can further decay to the biexcitonic band edge with a timescale $\hbar\gamma^{-1}$ assumed to be independent of the number of charge carriers. The decay of the exciton/biexciton from the corresponding band edge occurs on much longer timescales and is not described here (10).

In this picture, MEG will become efficient and may lead to CM when the timescale $\hbar\Gamma_S^{-1}$ is significantly shorter than the timescale $\hbar\gamma^{-1}$ associated with the relaxation of the initial exciton by other means. Furthermore, MEG can only occur at energies for which the initial excitation is at least twice above the material's band gap, $E_g$ to meet energy conservation. If excitation at energies twice above the band gap will result in 100% conversion to the biexcitonic state, then in principle, the energy efficiency of solar cell utilizing this process can exceed the SQ limit and reach values of 45% (11). Thus, materials which exhibit large CM efficiencies require that $\hbar\Gamma_S^{-1} << \hbar\gamma^{-1}$ for all exciton energies above $2E_g$.

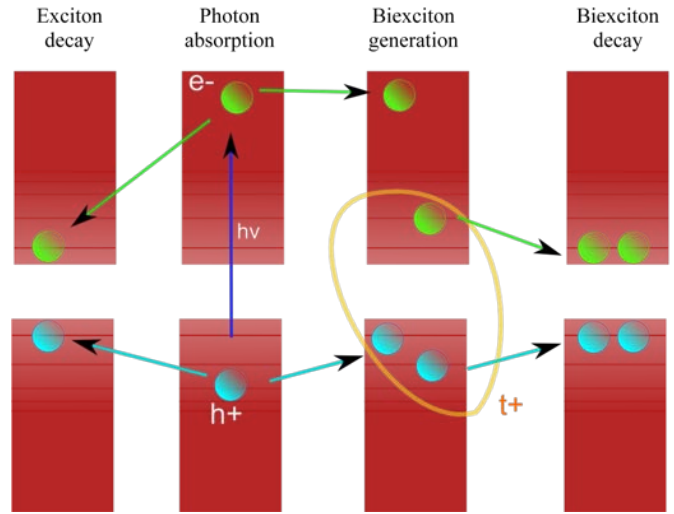

Figure 1: A sketch of the mechanism for biexciton generation in nanostructures. After absorption of a photon at time $t = 0$ an exciton is formed. This exciton can decay to the band edge with a timescale of $\hbar\gamma^{-1}$ typical to phonon emission or relax/transform to a resonant biexcitonic state within a typical timescale $\hbar\Gamma_S^{-1}$, which can then relax to the biexciton band edge on timescales similar to $\hbar\gamma^{-1}$. In the present example, the excited hole $h^+$ decays to a positive three-particle entity called a positive trion $T^+$.

The MEG phenomenon is known to occur in bulk semiconductors and has been studied for nearly 50 years (12). Strict selection rules and other competing processes in the bulk allow generation of multiexcitons at energies of $n \times E_g$ where $E_g$ is the band gap and $n > 3$, however, truly efficient MEG is observed only for $n > 5$ (13,14). In semiconducting nanocrystals (NCs) it was suggested that quantum confinement effects are important (7), enlarging Coulomb coupling and enabling a "phonon bottleneck" phenomenon that reduces the rates of electronic excitation decay. This engendered the concept that MEG in NCs may be efficient at lower values of $n$ (typically 2 to 3) (7). Indeed, MEG in semiconducting NCs has been reported recently for several systems (8,15-23), showing that the threshold is size and bandgap independent (16,17,22,23). However, more recent studies have questioned the efficiency of MEG in semiconducting NCs, in particular for CdSe (24) and InAs (25).

This controversy calls for theoretical assessment of the processes of MEG in nanostructures. In recent years several different theoretical treatments have been proposed (8,26-34) to address the efficiency of MEG in NCs. These can be classified to two groups: (a) Direct/Indirect absorption into the biexcitonic manifold (8,31) and (b) coherent/incoherent impact excitation (26-31,34,35). The purpose of the present review is to present a unified theory to calculate the efficiency of MEG and to derive the former approaches as approximations to the unified framework. We will argue that one approach based on the concept of incoherent impact excitation is the most suitable for MEG in semiconducting NCs. Our calculations support recent experiments on various systems reporting low efficiencies of $< 20\%$ at exciton energies near $3E_g$ (24,25,34,36,37).

## II. THEORY

Several different theoretical approaches have been suggested to describe MEG in semiconducting NCs. They can be classified to direct/indirect absorption (8,31) and coherent/incoherent impact ionization (26-31,34,35). In this section we will derive a unified theoretical approach to MEG based on the Green's function formalism and show how the different treatments emerge as approximations to the proposed framework. Within the unified approach, we will compute the total number of excitons generated when photon is absorbed as

$$n_{ex}(\omega) = \frac{r_S(\omega) + 2r_B(\omega)}{r_S(\omega) + r_B(\omega)} \quad (1.1)$$

where $r_S(\omega)$ is the rate of photon absorption into a single exciton manifold, $r_B(\omega)$ is the rate of photon absorption into a biexciton manifold, and $r(\omega) = r_S(\omega) + r_B(\omega)$ is the total photon absorption rate. We now present the theory for the photon absorption rate. Explicit expressions for $r_S(\omega)$ and $r_B(\omega)$ will be considered when discussing various approximations to equation (1.1).

### A. Green's function approach to MEG

We first describe the electronic structure of the NC in terms of the exciton Hilbert space. In this space, let $|0\rangle$ be the "Hartree-Fock" (HF) ground state, described as a Slater-determinant wavefunction. The single exciton states are: $\left|S_{i\sigma}^{a\sigma}\right\rangle = a_{a\sigma}^\dagger a_{i\sigma} |0\rangle$ where $a_{i\sigma}$ ($a_{a\sigma}^\dagger$) are annihilation (creation) operators for electron with spin $\sigma = \uparrow, \downarrow$ and single particle states $i$ ($a$). In this notation, the biexciton states are $\left|B_{j\sigma k\sigma'}^{c\sigma b\sigma'}\right\rangle = a_{b\sigma'}^\dagger a_{c\sigma}^\dagger a_{j\sigma} a_{k\sigma'} |0\rangle$. In what follows indices $i,j,k,l,$ are occupied (hole) state indices, $a,b,c,d$ unoccupied states (electron) and $r,s,t,u$ are general indices.

One can partition the electronic Hamiltonian as follows:

$$H = H_0 + H_{ph} - \mu\mathcal{E}\sin\omega t \quad (1.2)$$

Here,

$$H_0 = \begin{pmatrix} E_0 & 0 & W_{0B} & \cdots \\ 0 & H_S & W_{SB} & \cdots \\ W_{0B}^\dagger & W_{SB}^\dagger & H_B & \cdots \\ \vdots & \vdots & \vdots & \ddots \end{pmatrix} \quad (1.3)$$

is the unperturbed Hamiltonian with $E_0$ the HF ground-state energy, $H_S$ ($H_B$) the block-matrix containing matrix elements between singly excited (doubly excited) Slater wave functions. $W_{0B}$ is the block matrix describing the coupling elements of the HF ground state to the biexciton space, which will be neglected in the subsequent developments. There is no coupling of the HF ground state to single excitons ($W_{0S} = 0$). $W_{SB}$ is the matrix block describing the coupling between the excitons and biexcitons with the following non-zero matrix elements:

$$\begin{aligned}\left\langle S_{i\uparrow}^{a\uparrow}\middle|W\middle|B_{k\uparrow j\uparrow}^{b\uparrow c\uparrow}\right\rangle &= \delta_{ac}\left(V_{jikb} - V_{kijb}\right) + \delta_{ab}\left(V_{kijc} - V_{jikc}\right) \\ &\quad + \delta_{ij}\left(V_{kcab} - V_{ackb}\right) + \delta_{ki}\left(V_{acjb} - V_{jcab}\right) \\ \left\langle S_{i\uparrow}^{a\uparrow}\middle|W\middle|B_{k\uparrow j\downarrow}^{b\uparrow c\downarrow}\right\rangle &= \delta_{ab}V_{kijc} - \delta_{ki}V_{jcab} \\ \left\langle S_{i\uparrow}^{a\uparrow}\middle|W\middle|B_{k\downarrow j\uparrow}^{b\downarrow c\uparrow}\right\rangle &= \delta_{ac}V_{jikb} - \delta_{ji}V_{kbac}. \end{aligned} \quad (1.4)$$

The single exciton matrix $H_S$ includes diagonal terms given by $E_S \equiv \epsilon_a - \epsilon_i$ ($\epsilon_i$ is the single particle energy level) and off-diagonal terms given by:

$$\begin{aligned}\left\langle S_{a\uparrow}^{i\uparrow}\middle|W\middle|S_{b\uparrow}^{j\uparrow}\right\rangle &= V_{jbai} - V_{abji} \\ \left\langle S_{a\uparrow}^{i\uparrow}\middle|W\middle|S_{b\downarrow}^{j\downarrow}\right\rangle &= V_{jbai}\end{aligned} \quad (1.5)$$

In the above equations the Coulomb matrix elements are:

$$V_{rsut} = \iint d^3r\, d^3r' \left[\psi_r(\mathbf{r})\psi_s(\mathbf{r})\psi_u(\mathbf{r}')\psi_t(\mathbf{r}')/\epsilon|\mathbf{r}-\mathbf{r}'|\right], \quad (1.6)$$

$\epsilon$ is the dielectric constant of the NC estimated from Ref.

(38) for CdSe, Ref. (39) for InAs, and Ref. (40) for silicon. $\psi_r$ is the single electron wave function with energy $\epsilon_r$. From the above expressions, we find that an exciton $S_i^a$ can only couple to biexcitons $B_{ij}^{bc}$ or $B_{jk}^{ab}$. In the first case the electron $a$ decays into a negative trion $T_j^{bc}$ and in the second the hole $i$ decays to a positive trion $T_{jk}^{b}$ (27,30). Similar expressions hold for an exciton with spin down.

The dipole in Eq. (1.2) is given by the block matrix (indexed in a similar manner as $H_0$):

$$\mu = \begin{pmatrix} 0 & \mu_{0S} & 0 & \cdots \\ \mu_{0S}^\dagger & \mu_S & \mu_{SB} & \cdots \\ 0 & \mu_{SB}^\dagger & \mu_B & \cdots \\ \vdots & \vdots & \vdots & \ddots \end{pmatrix}, \quad (1.7)$$

and $\mathcal{E} \sin \omega t$ is the time-dependent electric field representing the interaction with the laser. $H_{ph}$ in Eq. (1.2) represents the phonon Hamiltonian and its coupling to the electronic degrees of freedom. Phonons will be incorporated phenomenologically below and $H_{ph}$ is not given explicitly.

To describe the process of MEG one must solve the dynamics generated by the above time-dependent Hamiltonian and calculate the projection onto the biexciton subspace. In practice, this is impossible due to the complexity of the many-body quantum dynamics. However, the electric field used in the relevant experiments is weak and a lowest order perturbative treatment with respect to the electric field is appropriate. The rate of photon absorption is given by the well-known golden rule formula assuming that at $t=0$ the system is in the ground HF state:

$$r(\omega) = -\frac{2}{\hbar} \mathcal{E}^2 Im \langle 0 | \mu G(\hbar\omega + E_0) \mu | 0 \rangle, \quad (1.8)$$

where $G(E) = (E - H_0 - \Sigma_{e-ph}(E))^{-1}$ is the Green's function of the unperturbed Hamiltonian and $\Sigma_{e-ph}(E) = -i\gamma/2$ is the phonon self-energy taken in the wide band limit (41), i.e., the real part of $\Sigma_{e-ph}(E)$ is negligible and the imaginary part is assumed to be a constant independent of energy. It represents the broadening of electronic states due to the coupling to phonons.

It is evident from Eq. (1.7) that the dipole operator couples the ground state only to the single exciton manifold. Therefore, the rate of photon absorption can be expressed as follows:

$$r(\omega) = -\frac{2}{\hbar} \mathcal{E}^2 Im \, \text{Tr}_S \left( \mu_{0S} G_S (E_0 + \hbar\omega) \mu_{S0} \right), \quad (1.9)$$

where

$$G_S(E) = \frac{1}{E - H_S - \Sigma_{SB}(E) + i\gamma/2} \quad (1.10)$$

is the single exciton block of $G(E)$ given in terms of the single exciton Hamiltonian $H_S$ and the self energy representing the coupling to the biexcitonic manifold:

$$\Sigma_{SB}(E) = W_{SB} (E - H_B + i\eta)^{-1} W_{BS}, \quad (1.11)$$

where $\eta$ is a positive infinitesimal number. The self energy is a matrix within the single exciton space.

Eqs. (1.9)-(1.11) provide a framework to compute the rate of absorption with the essential approximations being the perturbative treatment of the electric field, the wide band limit used to describe the coupling to phonons, and neglecting the contribution of triple excitonic states and higher to the single exciton self-energy. The density matrix formalism developed by Efros and coworkers (27) provides an alternative description within the same level of approximations. However, in the applications reported in Ref. (27) only several excitonic and biexcitonic states where included in the formulation, while the present approach accounts for all single- and bi-excitonic states. As will become clear below, the inclusion of the entire manifold of states is important, as was discussed recently in Ref. (33). In what follows, we will use Eqs. (1.9)-(1.11) as a starting point to derive different working approximations often used to express the efficiency of MEG in nanostructures.

### B. Weak Coulomb coupling limit

If the Coulomb coupling is treated within perturbation theory, the Green's function in Eq. (1.10) can be approximated by:

$$\begin{aligned} G_S(E) \approx & \, G_{S,\gamma}^{(2)}(E) + \\ & G_{S,\gamma}^{(0)}(E) W_{SB} G_B^{(0)}(E) W_{BS} G_{S,\gamma}^{(0)}(E), \end{aligned} \quad (1.12)$$

where $G_{S,\gamma}^{(0)}(E) = \left(E - H_S^{(0)} + i\gamma/2\right)^{-1}$ is the zero order Green's function of the single exciton Hamiltonian including the phonon self energy and $H_S^{(0)}$ is the diagonal part of $H_S$. $G_{S,\gamma}^{(2)}(E) = G_{S,\gamma}^{(0)}(E) \left(1 + W_S G_{S,\gamma}^{(0)}(E) \left(1 + W_S G_{S,\gamma}^{(0)}(E)\right)\right)$ is the Green's function of the single exciton Hamiltonian including the phonon self energy to second order in the Coulomb coupling, which is perturbative result of the Bethe-Salpeter treatment. $G_B^{(0)}(E) = \frac{1}{E - H_B^{(0)} + i\eta}$ is the zero order Green's function of the biexciton, where $H_B^{(0)}$ is the diagonal part of $H_B$ with terms $E_B \equiv \varepsilon_c - \varepsilon_j + \varepsilon_b - \varepsilon_k$.

The rate of absorption becomes a sum of two terms, $r(\omega) = r_S(\omega) + r_B(\omega)$, where:

$$r_S(\omega) = -\frac{2}{\hbar}\mathcal{E}^2 \text{Im } tr_S\left(\mu_{0S} G^{(2)}_{S,\gamma}(E_0 + \hbar\omega)\mu_{S0}\right)$$

$$r_B(\omega) = -\frac{2}{\hbar}\mathcal{E}^2 \text{Im } tr_S\left(\mu_{0S} G^{(0)}_{S,\gamma}(E_0 + \hbar\omega) W_{SB}\right.$$
$$\left. G^{(0)}_B(E_0 + \hbar\omega) W_{BS} G^{(0)}_{S,\gamma}(E_0 + \hbar\omega)\mu_{S0}\right) \quad (1.13)$$

To lowest order in the Coulomb coupling the rate $r_B(\omega)$ can also be written as:

$$r_B(\omega) = -\frac{2}{\hbar}\mathcal{E}^2 \text{Im}\sum_B \frac{F_B(E_0 + \hbar\omega)^2}{E_0 + \hbar\omega - E_B + i\eta}, \quad (1.14)$$

where

$$F_B(E) = \sum_S \frac{\mu_{0S} W_{SB}}{E - E_S + i\gamma/2}. \quad (1.15)$$

### C. Indirect absorption limit

Eq. (1.13) is the rigorous limit of the current approach to second order in the Coulomb coupling. To establish a connection with the indirect absorption approach (8), several additional approximations need to be introduced. As will become apparent shortly, these additional approximations are well defined but not always justified. The first approximation ignores the role of phonons, implying that one has to take the limit $\gamma \to 0$. Thus, $F_B(E)$ in Eq. (1.15) can be written as:

$$F_B(E) = \sum_S \frac{\mu_{0S} W_{SB}}{E - E_S} - i\pi \sum_S \mu_{0S} W_{SB} \delta(E - E_S) \quad (1.16)$$

The second approximation assumes that the imaginary part of $F_B(E)$ is negligible, consistent with the picture of virtual single-exciton states (8). After some simple algebra, one arrives at the expression for $r_B(\omega)$:

$$r_B(\omega) = \frac{2\pi}{\hbar}\mathcal{E}^2 \sum_B \left|\sum_S \frac{\mu_{0S} W_{SB}}{\hbar\omega + E_0 - E_S}\right|^2 \delta(\hbar\omega + E_0 - E_B),$$
(1.17)

which is the indirect absorption expression derived by Klimov and coworkers (8). In practice, one also has to replace $G^{(2)}_{S,\gamma}(E_0 + \hbar\omega)$ in Eq. (1.13) by its zeroth order limit $G^{(0)}_{S,\gamma}(E_0 + \hbar\omega)$, leading to:

$$r_S(\omega) = \frac{2\pi}{\hbar}\mathcal{E}^2 \sum_S |\mu_{0S}|^2 \delta(\hbar\omega + E_0 - E_s). \quad (1.18)$$

The above equations are based on two major assumptions in addition to the weak coupling limit. The first ignores the role of phonons. The second assumes that direct absorption into the single exciton manifold is negligible and thus the imaginary part of the $F_B(E)$ is neglected. Both seem to be unjustified from a physical point of view. In fact, when $E_B \to E_S$

Eq. (1.17) diverges as a result of the fact that boarding of the levels due to electron-phonon coupling and due to electron-electron coupling are ignored. This results in an infinite rate of absorption into the biexcitonic states leading to unphysical estimation of the MEG efficiency.

### D. Semi wide band approximation

The semi wide band limit is an approximation developed for practical reasons. It is quite difficult to obtain a full solution to Eqs. (1.9)-(1.11). A significant simplification can be achieved if one assumes that $\langle S|\Sigma_{SB}(E)|S'\rangle \approx -\frac{i}{2}\Gamma_S \delta_{S'S}$ is purely imaginary diagonal matrix with elements given by:

$$\Gamma_S = 2\pi\sum_B |W_{SB}|^2 \delta(E - E_B) \quad (1.19)$$

This expression can be derived from Eq.(1.11), where we have neglected the Coulomb couplings between biexcitons. In principle, these couplings can be included, but for practical reasons are ignored. Further neglecting the Coulomb coupling between excitons, i.e. replacing $H_S \to H^{(0)}_S$ in Eqs. (1.9) and (1.10), the rate of photon absorption is then given by:

$$r(\omega) = \frac{\mathcal{E}^2}{\hbar}\sum_S \frac{|\mu_{0S}|^2 (\gamma + \Gamma_S)}{(E_0 + \hbar\omega - E_S)^2 + (\gamma + \Gamma_S)^2/4}. \quad (1.20)$$

This allows for a natural definition of the photon absorption rates into single- and bi-excitonic states as follows:

$$r_S(\omega) = \frac{\mathcal{E}^2}{\hbar}\sum_S \frac{\gamma |\mu_{0S}|^2}{(E_0 + \hbar\omega - E_S)^2 + (\gamma + \Gamma_S)^2/4}$$

$$r_B(\omega) = \frac{\mathcal{E}^2}{\hbar}\sum_S \frac{\Gamma_S |\mu_{0S}|^2}{(E_0 + \hbar\omega - E_S)^2 + (\gamma + \Gamma_S)^2/4} \quad (1.21)$$

the number of excitons generated is then given by (Eq. (1.1)):

$$n_{ex}(\omega) = \frac{\displaystyle\sum_S \frac{(\gamma + 2\Gamma_S)|\mu_{0S}|^2}{(E_0 + \hbar\omega - E_S)^2 + (\gamma + \Gamma_S)^2/4}}{\displaystyle\sum_S \frac{(\gamma + \Gamma_S)|\mu_{0S}|^2}{(E_0 + \hbar\omega - E_S)^2 + (\gamma + \Gamma_S)^2/4}} \quad (1.22)$$

The above expression satisfies the limits $\lim_{\gamma \to 0} n_{ex}(\omega) = 2$ and $\lim_{\Gamma_S \to 0} n_{ex}(\omega) = 1$. It provides a convenient framework to calculate the efficiency of multiexciton generation beyond the perturbative treatment in the Coulomb coupling and it incorporates phonon effects on equal footing. Furthermore, Eq. (1.22) resembled the master equation result based on the perturbative treatment of incoherent impact ionization developed by us (30). In subsequent section we show numerically that the two approaches are, indeed, in excellent agreement.

## III. RESULTS AND DISCUSSION

### A. Density of single- and bi- exciton states

The calculation of the MEG efficiencies as defined in section II, by solving directly Eqs. (1.9)-(1.11) or by referring to one of the approximations, e.g., weak coupling limit, indirect absorption, and semi wide band limit, requires as input the screened Coulomb matrix elements between single and biexcitonic states $W_{SB}$. In some cases these Coulomb matrix elements between single excitonic states $W_S$ is also required. Therefore, one has to specify a framework which provides an accurate account of the electronic structure of the NC. The approach we adopt here is based on an atomistic semiempirical pseudopotential method that captures realistically the density of electronic states and provides a convenient framework to calculate the Coulomb matrix elements even at energies high above the band gap (42,43). This approach allows us to study the effect of NCs size (up to a diameter of ~3 nm and ~2000 electrons), photon energy (up to $3E_g$) and composition. The local screened pseudopotentials used in the results shown here were fitted to reproduce the experimental bulk band-gap and effective masses for CdSe (43), InAs (39), and silicon (44,45) neglecting spin orbit coupling (46). Furthermore, ligand potentials or hydrogen atoms (44,45) were used to represent the passivation layer (43). The resulting single-particle Schrödinger equation was solved in real space by the filter-diagonalization (FD) technique (47,48). FD allows construction of an eigensubspace of all energy levels up to $2E_g$ above the conduction band minimum (in principle, FD can be used to extract the energy level to any desired energy, but in practice for large NCs this is computational too demanding). From the FD solution, the density of states (DOS) was calculated by energy binning. As a check on the FD we also employed an alternative Monte Carlo method (49) which computes directly the DOS as $\pi^{-1}\,\mathrm{Im}\,Tr\left[(E - H + i\eta)^{-1}\right]$ (for the results shown, $\eta = 0.1\,\mathrm{eV}$). Using binning or self convolutions of the DOS, the exciton (DOSX) and biexciton (DOSXX) density of states (26) can be determined and are shown in Figure 2.

There are several features of DOSX and DOSXX which are turn out to play a major role in the process of MEG. The excitonic threshold occurs by definition at $E = E_g$. The two methods of calculating the DOSX agree well, indicating that the FD method is well converged and all states are generated within the energy window up to $2E_g$ above the conduction band minimum. This is required in order to obtain converged results for the Coulomb matrix elements. The biexcitonic threshold is $2E_g$. For higher energies the DOSXX grows with energy at a considerably faster rate than the DOSX (26), overtaking it at scaled energies which only slightly depend on the size and composition of the NC (between 2.3 and 2.5 $E_g$).

The crossover between DOSX and DOSXX and the magnitude of DOSXX (can approach $10^6\,\mathrm{eV}^{-1}$) implies that above the crossover energy, the biexcitonic manifold can be considered as a sink bath (50). Single excitons that decay to a biexcitonic manifold will remain there and recurrences are not likely to occur (33).

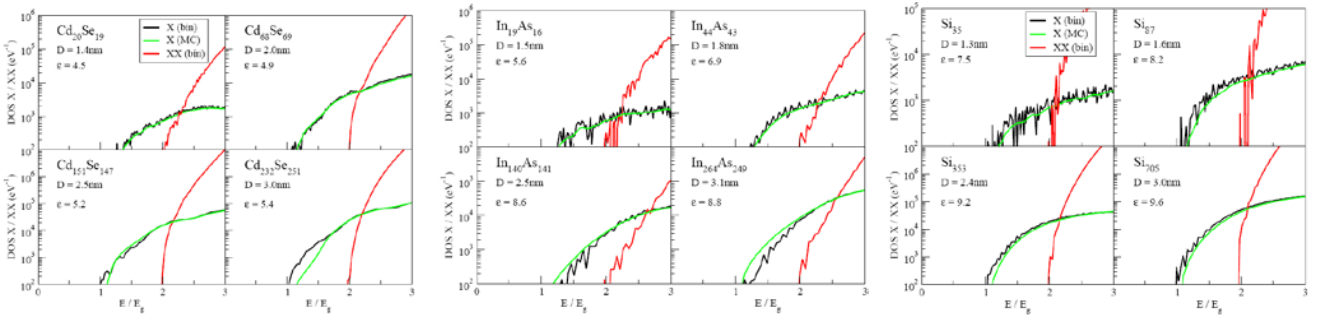

Figure 2: The single exciton (DOSX) and biexciton (DOSXX) density of states in various CdSe (left panels), InAs (middle panels), and silicon (right panels) NCs.

### B. Efficiency of MEG: the validity of the incoherent approach

In Figure 3 we plot the efficiencies of MEG for three prototype NCs: CdSe (II-VI), InAs (III-V) and Silicon (indirect band gap material). The details of the calculation will be outlined in subsection III B. The dashed curves in Figure 3 are the results obtained within the semi wide band limit (cf., Eq. (1.22)) and the solid curves are the results of the master equation approach where the number of excitons at steady states is determined by (30):

$$n_{ex}(\omega) = \sum_S p_S(\omega) \frac{2\Gamma_S + \gamma}{\Gamma_S + \gamma}, \quad (1.23)$$

where $p_S(\omega)$ is the absorption probability of generating an exciton:

$$p_S(\omega) = \frac{|\mu_{0S}|^2}{\sum_S |\mu_{0S}|^2}. \quad (1.24)$$

$\Gamma_S$ and $\mu_{0S}$ are given by Eq. (1.19) and (1.7), respectively,

and $\gamma / \hbar \approx 3 ps^{-1}$ is the value used for the phonon self-energy (51-55), typical of relaxation of the first excitonic states by electron-phonon coupling. Using a constant value for $\gamma$ is likely to give an upper bound to the MEG efficiency, since $\gamma$ increases with exciton energy, an effect not taken into account.

We observe excellent agreement between the master equation approach and the semi wide band limit when the density of states is high enough, e.g., at high energies or for large NCs. This agreement is not surprising given the similarity between the final expressions for the efficiencies based on the master equations approach (Eqs. (1.23)-(1.24) ) and the semi wide band limit (Eq. (1.22)). However, this is not entirely anticipated since the two approaches were derived based a completely different logic. The semi wide band limit assumes a coherent absorption into the biexcitonic manifold while the master equation does not. Furthermore, in the semi wide band limit, resonances decay exponentially in time, giving rise to a Lorentzian broadening, while in the master equation approach we have used a window function to represent resonances within the golden rule formula. Finally, the master equation is based on a perturbation treatment in the Coulomb coupling while the semi wide band limit includes all orders in $W_{SB}$. The agreement between the final two expressions indicates that coherent effects are not significant and that for the systems discussed herein, the Coulomb coupling is weak.

In both cases, MEG should become efficient when $\Gamma_S > \gamma$ as is the case for the smallest NCs. For larger NCs the efficiency of MEG decreases significantly in InAs and somewhat less so in CdSe and silicon. This is consistent with known results for bulk and is in agreement with recent experimental results on CdSe (24) and InAs (25) for NCs of diameter $D \geq 5 \mathrm{nm}$. We find that the efficiency of MEG decreases with nanoparticles size at a scaled energy (left panels of Figure 3) but increase with size at an absolute energy (right panels of Figure 3). As pointed out recently by Nozik and coworkers, for solar cell applications, the scaled energy is the proper representation (11). Thus, we conclude that MEG indeed becomes more efficient for confined systems and increases with decreasing nanocryatlline size.

The efficiencies of MEG calculated for the different systems studied herein are very similar. In this respect InAs shows distinct features compared to CdSe and silicon. For large particles, the efficiency of MEG drops to nearly 0% while for CdSe and silicon this is not the case. The behavior of InAs can be traced to the fact that it in a very narrow band gap material with light electron effective mass. Thus, at energies below $3E_g$, probing relatively low absolute energies for increasing NC's size, the density of single and biexcitonic states is quite small and the Coulomb coupling reduces as $R^{-3}$ (see Figure 5 below), giving rise to small values for $\Gamma_S$ and low MEG efficiencies.

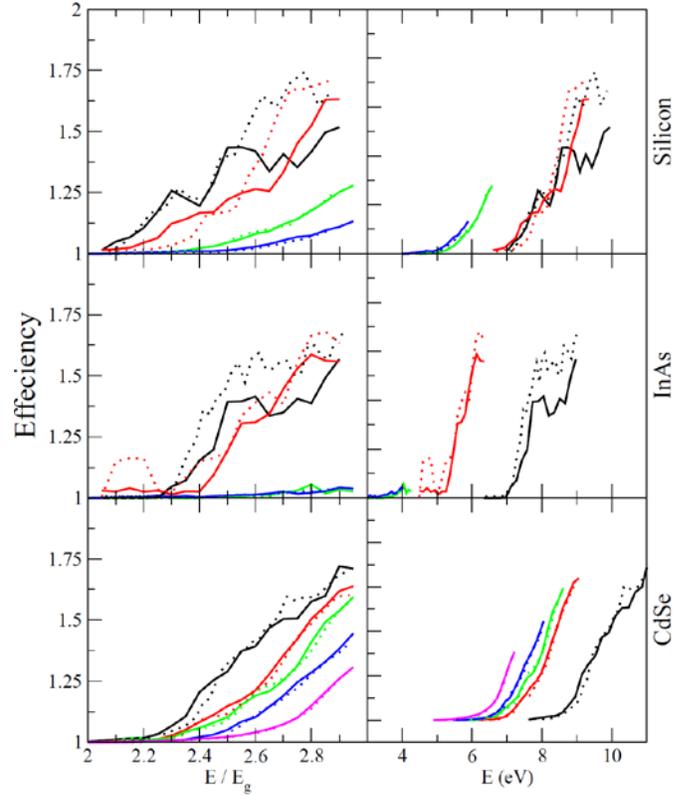

Figure 3: Multiexciton efficiencies calculated for CdSe (lower panels), InAs (middle panels) and silicon (upper panels) for several sizes. Left panels show efficiencies in scaled energy while the right panels are in absolute energy. Solid curves are the result based on the master equation with the rates calculated from Eq. (1.19) and the dashed curves are the results obtained from the semi wide band limit as described in subsection II.D. Systems sizes are (from black to magenta): $Cd_{20}Se_{19}$, $Cd_{68}Se_{69}$, $Cd_{83}Se_{81}$, $Cd_{151}Se_{147}$, and $Cd_{232}Se_{251}$; $In_{19}As_{16}$, $In_{44}As_{43}$, $In_{140}As_{141}$, and $In_{264}As_{249}$; $Si_{35}$, $Si_{87}$, $Si_{353}$, and $Si_{705}$.

### C. Scaling of the MEG process

In Figure 4 we show the values of $\Gamma_S$, $W_S$ (effective Coulomb coupling), and $\rho_T$ (trion density of state) for CdSe (left panels), InAs (middle plot), and silicon (right plots) at two NCs sizes. Each point in the figure represents an exciton $\left| S_{i\sigma}^{a\sigma} \right\rangle$ where the electron has an energy $\varepsilon_a$. The lowest panel depicts the density of trion states (DOTS). We only show the results for negative trions $\rho_S = \sum_{cbj} \delta\left(\varepsilon_a + \varepsilon_j - \varepsilon_b - \varepsilon_c\right)$. This is the density of states that enters Eq. (1.19) when selection rules (cf., Eq. (1.4)) are applied and the biexciton self-energy within the semi wide band limit reduces to (30):

$$\Gamma_S = 4\pi \sum_{cbj} \left| \left(V_{acjb} - V_{abjc}\right) \right|^2 \delta\left(\varepsilon_a - \left(\varepsilon_b + \varepsilon_c - \varepsilon_j\right)\right) \quad (1.25)$$

All the results shown in Figure 4 were obtained using a window representing the $\delta$-function of width 0.06 eV (results were not sensitive to widths above this value, to within a reasonable range).

There are several important features observed:

(a) In CdSe and silicon the DOTS increases at a given scaled energy as the size of the NC grows, reaching values above $10^1$ states per meV, justifying the incoherent treatment based on a master equation approach. In InAs this behavior is much weaker. At a given energy the DOTS of a NC increases with size, however, at a *scaled* energy, since $E_g$ *decreases* with size, the size dependence is weaker. In InAs the strong confinement makes $E_g$ highly sensitive to size causing a reduced sensitivity of the DOTS as a function of the scaled energy. Overall there are fewer trion states for InAs compared to CdSe and silicon because $E_g$ is smaller in the InAs and thus the absolute energy probed is lower.

(b) In the middle panels we show the effective Coulomb matrix element, defined as $W_S = \sqrt{\hbar \Gamma_S / (2\pi \rho_T)}$. The effective couplings are nearly energy independent, especially for large exciton densities, with a spread that decreases with NC size and spans 1-2 orders of magnitude. NCs of smaller diameter exhibit larger coupling elements. However, the coupling is not proportional to $D^{-1}$ as might be expected from a naïve analysis. In fact, $W_S$ scale as $D^{-3}$ for CdSe and InAs, as also discussed in Ref. (56); for silicon it scales with a slightly larger power due to the stronger dependence of the dielectric constant on the NC diameter. A log-log plots of $W_S$ is shown in Figure 5.

(c) The rate of exciton-biexciton transition at a given exciton energy spans 4-6 orders of magnitude, depending on the electron's energy. Since the effective Coulomb matrix elements are nearly independent of energy, the energy dependence observed for the rates reflect the behavior of DOTS. Thus, conclusions regarding the MEG process require the calculation of the rate for *all* excitons in a given energy, and may not be drawn from a limited arbitrary set. We find that due to quantum confinement, smaller NCs span a larger range of rates. In addition, smaller NCs have smaller DOTS but larger $W_S$. The net effect of combining the two quantities into the rate results in a larger rate for smaller NCs at a given scaled energy.

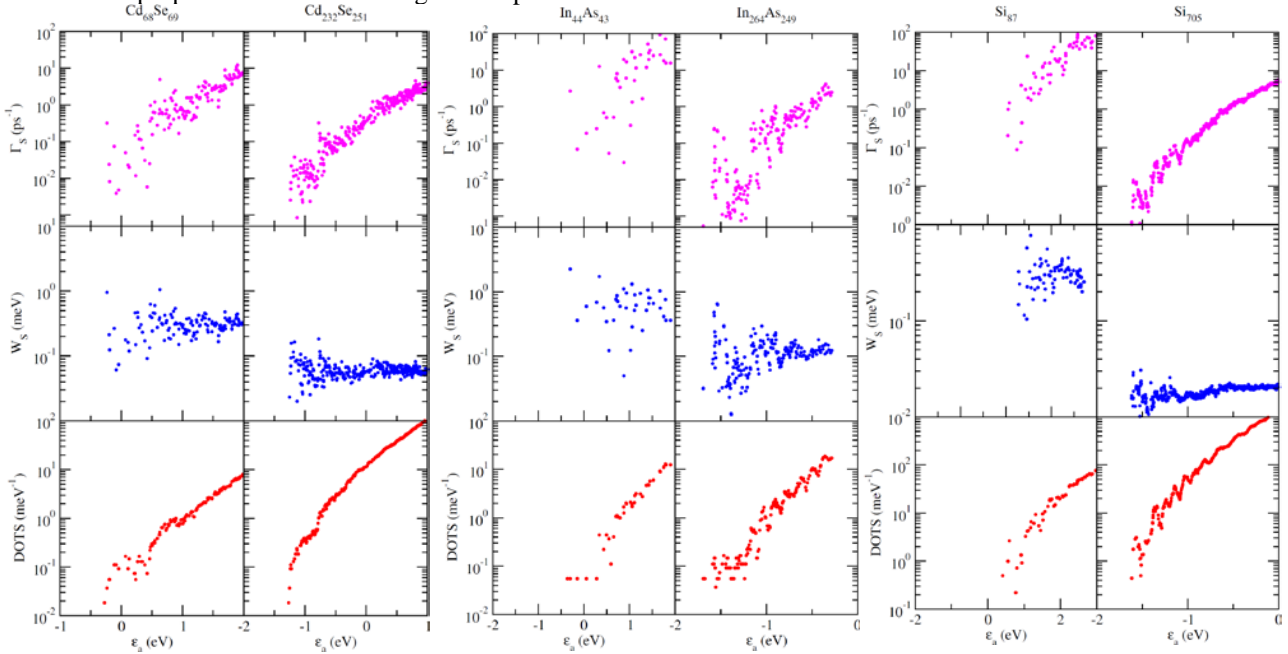

Figure 4: Plots of the density of trion states (DOTS, lower panels, red symbols), the average Coulomb coupling (middle panels, blue symbols) and the biexciton self-energy (Eq. (1.19), upper panels, magenta symbols) as a function of the energy of the electron (measured from the vacuum) for the three prototype nanocrystals at two system sizes: CdSe (left plot), InAs (middle plot), silicon (right plot).

## IV. SUMMARY

We have developed a unified approach to the treatment of MEG in nanocrystals. Our approach is based on the Green's function formalism, which in principle, leads to an exact description of MEG. It accounts for the screened Coulomb couplings between single- and bi-excitons, and between the exciton manifolds themselves. In addition, the formalism allows for the description electron-phonon couplings that are crucial for a complete description of MEG. Within this formalism, the efficiency of MEG is calculated from the rate of photon absorption.

In practice, the solution of the full Green's function formalism is difficult and several approximations need to be introduced. Common to all treatments, is the assumption that the phonon self-energy can be described within the wide band limit and the value of its imaginary part is taken from experimental work.

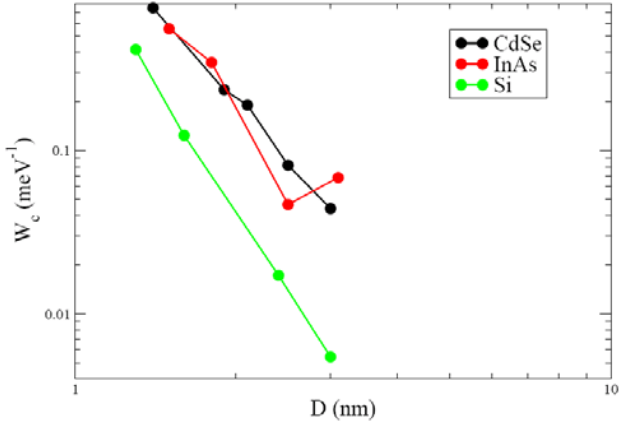

Figure 5: A log-log plots of the effective Coulomb coupling between single excitons and biexcitons as a function of the NC diameter for CdSe (black), InAs (red), and silicon (green).

The weak Coulomb coupling limit provides a solid framework when the matrix elements of $W$ are small compared to the phonon self-energy. Note that a systematic weak coupling treatment requires a consistent treatment of both the couplings between single- and bi-excitons and between the single excitons themselves, but not between the bi-excitons themselves (which leads to higher order contributions). From the weak coupling limit, we have derived the indirect absorption result of Klimov and coworkers (8). This required several additional assumptions, i.e., neglecting the direct absorption to single exciton states and neglecting the coupling to phonons. Both assumptions seem to be physically unjustified. This approach overestimates the efficiency of MEG as a result of singularities arising from resonances that are not broadened by electron-phonon or electron-electron couplings.

The second approach is based on the view that the Coulomb couplings between single- and bi-excitons are more important for describing MEG than those between single- and bi-excitons. Accordingly, the latter are neglected while the formers are treated to all orders within the semi wide band limit. The resulting expression for the photon absorption rate enables a natural dissection to the rate of absorption into single- and bi-excitons. The results of this approach were compared, theoretically and numerically, to those of previous work (30) which was based on an incoherent master equation approach. Excellent agreement between the two approaches indicates that MEG can be described as incoherent process.

We have also discussed the scaling of the MEG with respect to the size and composition of the nanocrystals, and with respect to the energy of the absorbed photon. The efficiency increases with energy, as expects, with an onset that is given by the crossover between DOSX and DOSXX. The effective Coulomb coupling is largely independent of energy, which implies that the scaling of the efficiency of MEG comes from the change in the density of states. In addition, at a scaled energy the efficiency of MEG increases with decreasing NC's size and is largely materials independent. At the NC sizes relevant for experiments, theory predicts that MEG efficiency is small, on the order of $< 20\%$, not sufficient to push the efficiency of thin film solar cells way above the SQ limit.

We would like to conclude with a positive view. Recent experiments by Gabor *et al.* (57) reported highly efficient generation of electron-hole pairs in single-walled carbon nanotube p-n junction photodiodes. Theoretical analysis of these experiments reveals the importance of the diode field along with the interplay between phonon emission, field acceleration, and multiexciton generation.(58) These results are encouraging and call for further investigation of the MEG in confined system.

**Acknowledgments.** We would like to thank Louis Brus and David Reichman for stimulating discussion and for helpful suggestions regarding this review. This research was supported by the Converging Technologies Program of The Israel Science Foundation (grant number 1704/07) and The Israel Science Foundation. We would like to thank the Center for Re-Defining Photovoltaic Efficiency Through Molecule Scale Control, an Energy Frontier Research Center funded by the U.S. Department of Energy, Office of Science, Office of Basic Energy Sciences under Award Number DE-SC0001085 for support.